# A Modified Interference Approximation scheme for Improving Preamble Based Channel Estimation Performance in FBMC System

Radwa A. Roshdy[1], Mohamed A. Aboul-Dahab[2] and Mohamed M. Fouad[3]

[1]Department of Electrical Engineering, Higher Technological Institute,
10th of Ramadan City, Egypt
[2]Department of Electronics and Communications Engineering,
Arab Academy for Science Technology and Martine transport, Cairo, Egypt
[3]Department of Electronics and Communications Engineering,
Zagazig University, Zagazig, Egypt

*ABSTRACT*

*Filter bank multicarrier (FBMC) is considered a competitive waveform candidate for 5G that can replace orthogonal frequency division multiplexing (OFDM). However, channel estimation (CE) is a big challenge in FBMC because it suffers from intrinsic interference which is due to the orthogonality of the subcarrier functions in the real field only. In this paper, we investigate a proposed modified interference approximation scheme (M-IAM) by approximating the intrinsic interference from the neighbouring pilots to accommodate the complex channel frequency and thus improving CE performance besides simplifying its processing. The M-IAM scheme has a larger pseudo pilot magnitude than other conventional preamble schemes, namely the interference approximation method (IAM) with its versions (IAM-C) and (E-IAM-C); in addition to the novel preamble design (NPS). In addition, the proposed (M-IAM) scheme is characterized by the lower transmitted power needed. The CE performance of the M-IAM is investigated through 512 and 2048 subcarriers via different types of outdoor and indoor multipath fading channels that are time-invariant such as IEEE 802.22, IEEE 802.11, Rician, and additive white Gaussian noise (AWGN), as well as time-varying channels such as Rayleigh and Vehicular A (Veh-A). Simulation results demonstrate that the proposed M-IAM scheme achieves a lower bit error rate (BER), lower normalized mean square error (NMSE) and lower peak-to-average power ratio (PAPR) over the conventional preamble schemes under the aforementioned channel models. The proposed scheme has the advantage of saving the transmitted power, a requirement that could match 5G low power requirements.*

## 1. INTRODUCTION

Orthogonal frequency division multiplexing (OFDM) is more efficient as a multicarrier modulation technique in 4G, thanks to its higher flexibility in supporting adaptive subcarrier modulation, its lower complexity implementation and its simpler channel estimation processing and equalization [1]. However, OFDM isn't an appropriate candidate to be a multicarrier modulation technique in 5G because of its limitations that hinder it to fulfil its requirements that aren't addressed by 4G, such as latency reduction (nearly 1 ms), higher capacity, higher data rates (up to 10 Gbps) and longer battery life [2]. OFDM requires high synchronization accuracy and any synchronization error destroys the orthogonality and causes inter-symbol interference (ISI). Moreover, it includes cyclic prefix (CP) which limits the spectral efficiency. OFDM has also long trip time, which makes it inefficient in machine to machine (M2M) applications. Also, it has high out of band (OOB) emission which results from the large power of spectrum side lobes [3]. Thus the need for a new multicarrier modulation technique to replace OFDM in 5G is essential. Filter bank multicarrier (FBMC) has many advantageous characteristics that make it a promising alternative multicarrier modulation over OFDM in 5G [3-5]. In this technique, CP is replaced by additional filtering at transmitter and receiver besides IFFT/FFT processes by filtering each output of the FFT using a frequency version of a low pass filter (LPF) termed as a prototype filter. The prototype filter is designed to be a finite impulse response to reduce the inter-symbol interference (ISI) and these results in increasing bandwidth efficiency [6]. The input to FBMC is divided into real and imaginary components and each component is multiplied to a phase shift pattern. Also FBMC is more flexible to exploit white spaces in cognitive radio networks and it is also more robust to synchronization and frequency misalignment [7]. However, FBMC is more sensitive to what is so-called intrinsic interference [3], which results due to the orthogonality of the subcarrier functions in the real field only. This means that this interference always exists between neighbouring subcarriers and symbols [8]. Although the offset quadrature amplitude modulation (OQAM) purpose is to dodge the intrinsic interference by staggering the real and imaginary elements on the time-frequency lattice [9], the OQAM cannot avoid the intrinsic interference in multi-path channel, because of the complex-valued response of the channel. Consequently, the current channel estimation methods of OFDM cannot be applied to FBMC. Channel estimation (CE) methods of FBMC attract numerous researches, especially in multipath fading channels [10]. FBMC CE methods include preamble-based [11]-[13] and scattered pilots-based techniques [11], [14], and [15] techniques. The former ones are supposed to cancel the undesired interference, like interference cancellation method (ICM) or effectively exploiting it to improve the estimation performance, like interference approximation method (IAM) [12]. IAM is a type of a preamble based CE which accommodates the complex channel frequency response (CFR) by constructing complex-pseudo pilots [16], [27].

The IAM method includes different types of schemes such as IAM-R, IAM-C, and E-IAM-C and all of these schemes aim at constructive approximating interference to improve channel estimation performance. IAM-R that has been proposed in [12] generates pseudo-pilots of large magnitude in a simpler way. All pseudo pilots are real with nulls at the first and third symbols of the preamble. IAM-C that has been proposed in [26]-[28] uses pseudo-pilots which are either purely real or imaginary at all the subcarriers. This is done by setting the middle FBMC symbol equal to that in IAM-R but with the pilots at the odd subcarriers multiplied by j (to make it imaginary) and hence the magnitude of the pseudo pilots is maximized. An extended method of IAM-C (E-IAM-C) with magnitude larger than that of IAM-C is given in [13] and is characterized by the best CE performance gain over the other IAM schemes. However, more power is required to be transmitted because all three pseudo pilots of this preamble are nonzero, unlike the previous schemes. In [17] a novel preamble design (NPS) has been proposed based on a combination between IAM and ICM. This approach has resulted in better CE performance than that of previous IAM and ICM schemes despite its smaller magnitude. In NPS, all of the three pseudo pilots are also nonzero and thus more transmitter power is required. The main problem with preamble schemes is constructing pseudo pilots with maximum magnitude and minimum peak-to-average power ratio (PAPR) [13], [17]. So, it is quite interesting to develop a modified IAM through which maximum magnitude and minimum PAPR can be optimized. A modified IAM that can achieve this goal has been previously introduced by the authors in [18].

In this paper, the proposed modified IAM (M-IAM) is analysed, and its performance is more investigated and compared to other preamble schemes under both time-invariant and time-varying channels. In addition to the bit error rate (BER) and peak-to-average power ratio (PAPR), the normalized mean square error (NMSE) performance is investigated.

The rest of the paper is organized as follows: In Section 2 the IAM channel estimation model is described. Section 3 reviews the conventional preamble schemes. A modified IAM is presented and discussed in Section 4. Simulation results, including BER, NMSE, and PAPR are reported in Section 5. Section 6 concludes the paper.

## 2. IAM CHANNEL ESTIMATION METHOD

In this model, the filter bank multicarrier (FBMC) is used. FBMC filters each subcarrier individually using a real symmetric prototype filter $g_{m,n}(l)$ where $(l)$ represents the time domain. The impulse response of the prototype filter is defined by $g(l - n\frac{M}{2})$ where $(m,n)$ being frequency-time (FT) points; $m$ is a subcarrier index and $n$ is a time index of OQAM symbol and $M$ is an even number of subcarriers [19]. The pulse g is designed so that the neighbouring subcarrier functions $g_{m,n}(l)$ are only orthogonal. This results in some ISI or/and ICI at the output of the analysis filter bank (AFB), which is purely imaginary and known as intrinsic interference, even in the absence of channel distortion and noise and with perfect time and frequency synchronization. The intrinsic interference is given in [13] as:

$$\sum g_{m,n}(l) g_{p,q}(l) = j \langle g \rangle_{m,n}^{p,q} \qquad (1)$$

Where $(p,q)$ is a set of FT points which interfere with $(m,n)$ FT points. $p$ is a subcarrier index and $q$ is an OQAM symbol time index. Assuming that the channel is approximately frequency flat at each subcarrier and constant for of the prototype filter [12], then FBMC analysis filter bank output at the $p_{th}$ subcarrier and $q_{th}$ FBMC symbol is given [8] as:

$$y_{p,q} = H_{p,q} d_{p,q} + j \sum_{m=0,(m,n)\neq(p,q)}^{M-1} \sum_{n} H_{m,n} d_{m,n} \langle g \rangle_{m,n}^{p,q} + \eta_{p,q} \qquad (2)$$

$$y_{p,q} = H_{p,q} d_{p,q} + I_{p,q} + \eta_{p,q} \qquad (3)$$

Where $d_{p,q}$ is a real OQAM symbol, $H_{p,q}$ a CFR at that FT points, $I_{p,q}$ is associated with interference and $\eta_{p,q}$ is associated noise component, which is considered Gaussian with zero mean and variance $\sigma^2$. A common assumption is that, with a good time-frequency localized pulse, contributions to $I_{p,q}$ only come from the first order neighbouring of, $(p,q)$, namely. $\Omega_{p,q} = \{(p, q \pm 1), (p \pm 1, q \pm 1), (p \pm 1, q)\}$.

If the CFR is almost constant over this neighbourhood, then (2) can be re-written as:

$$y_{p,q} = H_{p,q} C_{p,q} + \eta_{p,q} \qquad (4)$$

Where:

$$C_{p,q} = d_{p,q} + jv_{p,q} \tag{5}$$

Where $d_{p,q}$ is a real OQAM symbol and $v_{p,q}$ is a virtual transmitted symbol $(p,q)$. If the interference in $jv_{p,q}$ in (5) is resulting only because of the immediate neighbours of $(p,q)$ FT points and these FT points carry training symbols, then interference can be computed and approximated. This requires that all input symbols in the immediate neighbourhood of $(p,q)$ FT point to be known and this means that, for each subcarrier, at least 1.5 complex symbols are required for training symbols [8]. This can serve as a pseudo pilot to compute an estimation of the CFR at the corresponding FT points. From (4), CFR estimation will be:

$$\tilde{H}_{p,q} = \frac{y_{p,q}}{C_{p,q}} \approx H_{p,q} + \frac{\eta_{p,q}}{C_{p,q}} \tag{6}$$

$$\tilde{H}_{p,q} \approx H_{p,q} + \frac{\eta_{p,q}}{d_{p,q} + jv_{p,q}} \tag{7}$$

Despite a simple one-tap equalization process with a prototype filter $g$ that has good localization properties in time and frequency domains may be sufficient to restore the real orthogonality, this equalization requires complex-valued CE. From (7), the training symbols (pilots) should be chosen to minimize the effect of the noise component. This means that the greater of the power the term ($d_{p,q} + jv_{p,q}$) the better the estimation will be.

## 3. THE CONVENTIONAL PREAMBLE SCHEMES

One of the IAM schemes is the IAM-C; this scheme is named IAM-C because its training symbols are combined between real and imaginary values. In this scheme, nulls are placed at the first and third FBMC training symbols of the preamble is $d_{P,0} = d_{p,2} = 0$ for all values of subcarriers. Then the imaginary part of the pseudo pilot $(p,1)$ will come only from the symbols at the positions $(p \pm 1,1)$ with the pilots at the odd subcarriers multiplied by (j) [16]. This scheme is shown in Fig. 1(a), with OQPSK modulation and M = 8.

E-IAM-C is an extended version of IAM-C, in which the pseudo pilots' magnitude is greater than that of IAM-C. At an odd-indexed subcarrier $p$, with the pilot $\pm jd$ in the middle, $\mp d$ is placed at the right-hand side, and it is negative, $\pm d$ at the left hand side. The E-IAM-C pseudo-pilots magnitude is larger than that achieved by IAM-C and IAM-R [13]. An example of the E-IAM-C scheme with OQPSK modulation and M=8 is shown in Fig. 1(b).

The novel preamble structure, NPS, which is proposed in [17], is based on both IAM and ICM methods by using asymmetry patterns to get a combination between them. NPS has pseudo pilots with magnitude smaller than other conventional IAM schemes and larger than other ICM schemes. But take into account interference cancellation characteristics. NPS preamble scheme with OQPSK modulation and M=8 is shown in Fig.1(c).

## 4. A Proposed Modified IAM (M-IAM) Scheme

Although E-IAM-C and NPS preamble schemes improve channel estimation performance more than other preamble schemes such as IAM-R and IAM-C, they require more power to be transmitted. This is because all of the three FBMC pseudo pilots of E-IAM-C and NPS are nonzero. Since some applications like 5G, require low power consuming a novel preamble scheme is proposed, in which the transmitted power is made lower than that of the E-IAM-C and NPS schemes. According to [8] $C_{p,q}$ is defined as follows:

$$C_{p,q} = d_{p,q} + j \sum_{(m,n) \in \Omega_{p,q}} d_{m,n} \langle g \rangle_{m,n}^{p,q} \tag{8}$$

Where $\langle g \rangle_{m,n}^{p,q}$ are the interference weights for the neighbours ($m,n$) that are belonging to $\Omega_{p,q}$ of each FT point ($p,q$) interest.

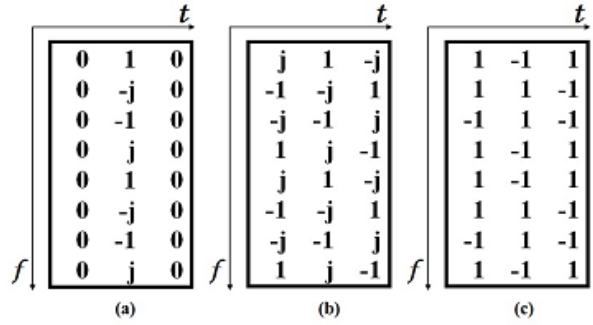

Figure 1. Preamble schemes for (a) IAM-C, (b) E-IAM-C, and (c) NPS.

This interference weights can be a priori computed based on the employed prototype filter g. These weights follow a specific pattern for any choice of g which can be written for all $q$ as:

$$\begin{matrix} (-1)^p \varepsilon & 0 & -(-1)^p \varepsilon \\ (-1)^p \delta & -\beta & (-1)^p \delta \\ -(-1)^p \gamma & d_{p,q} & (-1)^p \gamma \\ (-1)^p \delta & \beta & (-1)^p \delta \\ (-1)^p \varepsilon & 0 & -(-1)^p \varepsilon \end{matrix} \tag{9}$$

Where:

$$\beta = e^{-j(\frac{2\Pi}{M})(\frac{L_g-1}{2})} \sum_{l=0}^{L_g-1} g^2(l) e^{j\frac{2\Pi}{M}l} \tag{10}$$

$$\gamma = \sum_{l=\frac{M}{2}}^{L_g-1} g(l) g(l - \frac{M}{2}) \tag{11}$$

$$\delta = -je^{-j(\frac{2\Pi}{M})(\frac{l_g-1}{2})} \sum_{l=\frac{M}{2}}^{L_g-1} g(l) g(l - \frac{M}{2}) e^{j\frac{2\Pi}{M}l} \tag{12}$$

$$\varepsilon = e^{\mp j(\frac{2\Pi}{M})(\frac{L_g-1}{2})} \sum_{l=\frac{M}{2}}^{L_g-1} g(l)g(l-\frac{M}{2})e^{\pm j\frac{2\Pi}{M}l} \tag{13}$$

According to [8] and [14], the maximum values of interference weights are for $\gamma, \beta, \delta,$ and $\varepsilon$ respectively. So we send pseudo pilots in the position of $\gamma$ and $\beta$; and they have the maximum weights and the interference weight pattern will be:

$$\begin{matrix} 0 & -\beta & 0 \\ -(-1)^p \gamma & d_{p,q} & (-1)^p \gamma \\ 0 & \beta & 0 \end{matrix} \tag{14}$$

Nulls are placed at the position of $\delta$ and $\varepsilon$ thus transmitted power is reduced. We choose signs of the neighbouring pilots such that the pseudo-pilots' magnitude is maximized according to weight patterns in (9).

At an even-indexed subcarrier $p$, the middle pilots are real with $(-d)$ and $(+d)$ respectively. At the left and the right corresponding neighborhood positions, we placed nulls. At an odd-indexed subcarrier $p$, the middle pilots are also real with $(+d)$ and $(-d)$ respectively. At the left and the right corresponding neighborhood positions we place $(-d)$ and $(+d)$ for positive middle pilot $(+d)$ and nulls for negative middle pilots $(-d)$. From (14), the interference components, from the 1st and 3rd neighbors' symbols, cancel each other for the negative pilots of odd indexed and consequently have no contribution to the pseudo-pilot. We use nulls at the corners of the negative odd-indexed pilots and at the even-indexed pilots to reduce transmitted power lower than that of E-IAM-C and NPS. In the proposed M-IAM scheme, three preamble pilots aren't transmitted in each subcarrier as in E-IAM-C and NPS, only in the positive odd-indexed pilots of subcarrier and thus we save transmitted power. This of course leads to lower consumed energy during. All pilots in the M-IAM are real values and this makes it simpler than IAM-C and E-IAM-C which use imaginary pilots. The proposed preamble structure configuration for M = 8 and OQPSK modulation is shown in Fig. 2.

The pseudo pilot magnitude equals the power of $(d_{p,q} + jv_{p,q})$, which can be expressed as:

$$Mag = E\left\{ \left| \begin{matrix} d_{p,1} + j(d_{p+1,1}\langle g \rangle_{p+1,1}^{p,1} + d_{p-1,1}\langle g \rangle_{p-1,1}^{p,1} \\ + d_{1,q+1}\langle g \rangle_{1,q+1}^{p,1} + d_{1,q-1}\langle g \rangle_{1,q-1}^{p,1} \end{matrix} \right|^2 \right\} \tag{15}$$

We calculate the pseudo pilot magnitude by substitute (9) and the proposed M-IAM scheme into (15). At odd-indexed subcarrier the pseudo pilot magnitude is given by:

$$Mag = d\sqrt{1 + 4(\beta + \gamma)^2} \tag{16}$$

At even-indexed subcarrier the pseudo pilot magnitude is given by:

$$Mag = d\sqrt{1 + 4(\beta)^2} \tag{17}$$

Table 1 contains a comparison between the expressions of the pseudo magnitude of different types of preamble schemes. It is depicted that from Table I that the pseudo pilot magnitude of the proposed M-IAM scheme is larger than that of IAM-C and NPS at the odd indexed subcarriers. At even indexed subcarriers it is larger than that of NPS and smaller than that of IAM-C. Compared to E-IAM-C the proposed M-IAM has smaller pseudo pilot magnitude, this is because the proposed M-IAM transmit three preamble pilots at only positive odd indexed subcarriers and locate nulls at the corners of the negative odd-indexed pilots and at the even-indexed pilots ,unlike E-IAM-C that transmit three preamble pilots at both even and odd indexed subcarriers.
The proposed M-IAM scheme is simpler than the scheme of IAM-C and E-IAM-C as it doesn't contain imaginary training symbols which degrade the BER performance of IAM schemes as it is depicted from BER simulation results.

The algorithm that illustrates the M-IAM scheme was introduced in algorithm 1.

---

**Algorithm 1**: The modified IAM (M-IAM)

**Inputs**
M: no. of subcarriers
K: overlapping factor
$L_g$: prototype filter length (=KM)

**Process**
Calculate $\beta$ in equation (3.3)

Calculate $\gamma$ in equation (3.4)

Calculate $\delta$ in equation (3.5)

Calculate $\varepsilon$ in equation (3.6)

**IF** ( $\beta > \gamma$ ) **&&** ( $\gamma > \delta$ ) **&&** ( $\delta > \varepsilon$ )

Put nulls at the position of $\delta$ and $\varepsilon$ in equation (3.2)

Put ones at the position of $\beta$ and $\gamma$ in equation (3.2)

**END**

Alternate the sign of the middle pilot between positive and negative

**IF** the sign of the middle pilot is positive

Put nulls at the position of $\gamma$ at the negative middle pilot

**Else**

Put nulls at the position of $\gamma$ at the positive middle pilot

**END**

---

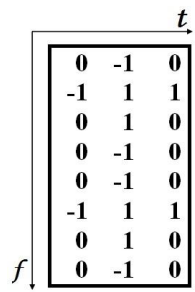

Figure 2. Modified IAM (M-IAM) scheme.

## 5. SIMULATION RESULTS

The performance of the proposed M-IAM is compared to the conventional preamble scheme through using the simulation results which were carried out through m-files using MATLAB package. The BER, NMSE, and PAPR of the proposed M-IAM versus IAM-C, E-IAM-C, and NPS are illustrated with different numbers of subcarriers of M=512 and M=2048. The simulation was carried out in outdoors and indoors multipath fading channels environment. The channel models selected are time-invariant, namely the IEEE 802.22, IEEE 802.11, rician and AWGN, as well as time-varying channel models, namely Rayleigh and Veh-A. They are also assumed to be frequency selective, multipath, fading channels. Channel Paths are illustrated in table 2. The table illustrates the channel profile of IEEE 802.22, IEEE 802.11, veh-A and Rayleigh channel. Rician and AWGN channels aren't included in the table as they don't have average powers or delays. The

Noise which is assumed Gaussian with zero mean and variance $\sigma^2$ is added at the receiver front-end. The FBMC system which is implemented in our simulation is FBMC spreading not FBMC with a polyphase network (PPN) despite the later system has an effective reduction in computational complexity. This is because the other conventional preamble schemes that are available in the literature have used FBMC spreading. So to get a fair comparison of our proposed M-IAM to other schemes we also us FFBMC spreading. In our, simulation the preambles are followed by pseudo-random data. The results considering BER of The aforementioned conventional preambles were verified as results in [17] in the case of IEEE 802.22 and Veh A with M= 2048. The results considering NMSE of E-IAM-C and IAM-C were verified as results in [13] in the case of Veh A with a number of subcarriers equals to 512, the results of our simulation aren't the same as those in [13] but it is very close. The Simulation parameters of the system were demonstrated in Table 2.

Table 1. Magnitude of Different Types of Conventional Preamble Schemes.

|  | IAM-C | E-IAM-C | NPS | Proposed M-IAM |
|---|---|---|---|---|
| **Odd indexed subcarriers** | $d\|1+2\beta\|$ | $d\|1+2(\beta+\gamma)\|$ | $d\sqrt{1+4(\beta+\delta)^2}$ | $d\sqrt{1+4(\beta+\gamma)^2}$ |
| **Even indexed subcarriers** | $d\|1+2\beta\|$ | $d\|1+2(\beta+\gamma)\|$ | $d\sqrt{1+4(\beta-\gamma)^2}$ | $d\sqrt{1+4\beta^2}$ |

Table 2. System Parameters.

| Parameter | Value |
|---|---|
| FFT/IFFT size of FBMC | 512, 2048 |
| Numbers of subcarriers (M) | 512, 2048 |
| Numbers of symbols/subcarriers | 40 |
| Bits/symbols | 2 |
| Channel models | Time invariant (IEEE 802.22, IEEE802.11, Rician, AWGN) and Time Variant (Rayleigh, Veh-A) with low frequency selectivity |
| Modulation | QPSK |
| Channel coding | Convolution code ($g_1$=133, $g_2$=171, code rate=0.5) |
| Type of prototype filter | Raised cosine filter with a roll off factor =1 |

| Overlap factor | 4 |
|---|---|
| Path Numbers | IEEE 802.11 ( From 1 to 21) |
| | IEEE 802.22 ( From 1 to 6) |
| | Veh-A (From 1 to 6) |
| | Rayleigh ( 1, 2) |

## 5.1. Bit Error Rate (BER) Comparison

Fig.3 to Fig. 8 show the BER performance of the proposed M-IAM scheme as compared to other preamble schemes, namely NPS, E-IAM-C and IAM-C for (a) M=512 and (b) M= 2048 in environments characterized by time-varying and time-invariant channels. Fig. 3 illustrates the BER performance of the different preamble schemes in the IEEE 802.22 channel. It is observed that the performance of the M-IAM has a gain of 2dB over the NPS at BER of 10-3 for M=512. However, the NPS has a slightly better performance in case The of M=2048. It is also observed that the BER performance of the proposed M-IAM is nearly the same for M=512 and M= 2048. The IAM-C has the worst BER performance because of the limiting effect of the SNR regime. Fig.4 illustrates the BER performance of the various preamble schemes in an IEEE 802.11 channel. The proposed M-IAM scheme has a gain of 3.5 dB over the nearest curve of the other schemes (which is that of NPS) at BER of 10-3 for M=512. In case The of M=2048, the performance of the proposed M-IAM is the same as that of NPS. The proposed M-IAM improves BER performance more than NPS, E-IAM-C, and IAM-C because they have inflexible and inconvenient preamble design, unlike the proposed M-IAM. It is quite interesting that the BER performance of the proposed M-IAM under these channel conditions hasn't varied much from that of the previous case the shown in Fig.3. In Fig.5, the BER performance of the various preamble schemes are illustrated in case the of a Rayleigh channel. The proposed M-IAM scheme has an SNR again of 2.5 dB at BER of 10-3 over the nearest scheme (NPS) in case of M=512. However, no again is nearly achieved at M=2048. It is also observed that the BER performance of the proposed M-IAM is nearly the same for M=512 and M= 2048. Fig. 6 illustrates the BER performance of the different preamble schemes in the rising channel. It is observed that the proposed M-IAM has an again 2 dB over the nearest curve of the other schemes (which is that of NPS) at BER of 10-3 for M=512. In case the of M=2048, the proposed M-IAM has an again 3 dB over the NPS scheme. It is also observed that the BER performance of the proposed M-IAM is nearly the same for M=512 and M= 2048. Fig. 7 illustrates the BER performance of the different preamble schemes in the Veh-A channel. It is observed that the proposed M-IAM has a gain 2 dB over the NPS at BER of 10-3 for M=512 and 3.5 dB in the case of M=2048. It is also observed that the BER performance of the proposed M-IAM for M=512 is better than the BER performance of the proposed M-IAM for M=2048 Fig. 8 illustrates the BER performance of the different preamble schemes in AWGN channel. The BER performance of the proposed M-IAM is the same as that of NPS for M=512 and M=2048.

## 5.2. Normalized Mean Square Error (NMSE) Comparison

The normalized mean square error is defined according to the equation (18) as following [20]:

$$NMSE = \frac{1}{M} \sum_{m=0}^{M} \frac{\left\| H - \tilde{H} \right\|^2}{\left\| H \right\|^2} \qquad (18)$$

Where $H$ the CFR and $\tilde{H}$ is the CFR estimation. Fig.9 to Fig. 13 show the NMSE performance of the proposed M-IAM scheme as compared to other preamble schemes, namely NPS, E-IAM-C and IAM-C for (a) M=512 and (b) M= 2048 in environments characterized by time-varying and time-invariant channels. It is depicted that the M-IAM improves the NMSE performance effectively; this is because the variance $\dfrac{\eta}{d_{p,q}+jv_{p,q}}$ is reduced with the proposed M-IAM.

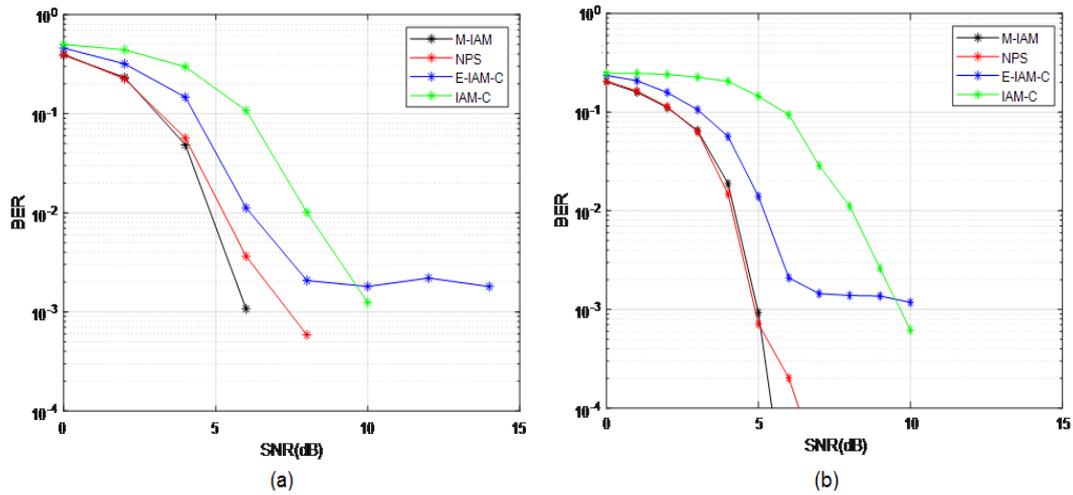

Figure 3. BER Performance for Different Preamble Schemes in IEEE 802.22 Chanel Models for M: (a) 512, (b) 2048 Subcarriers.

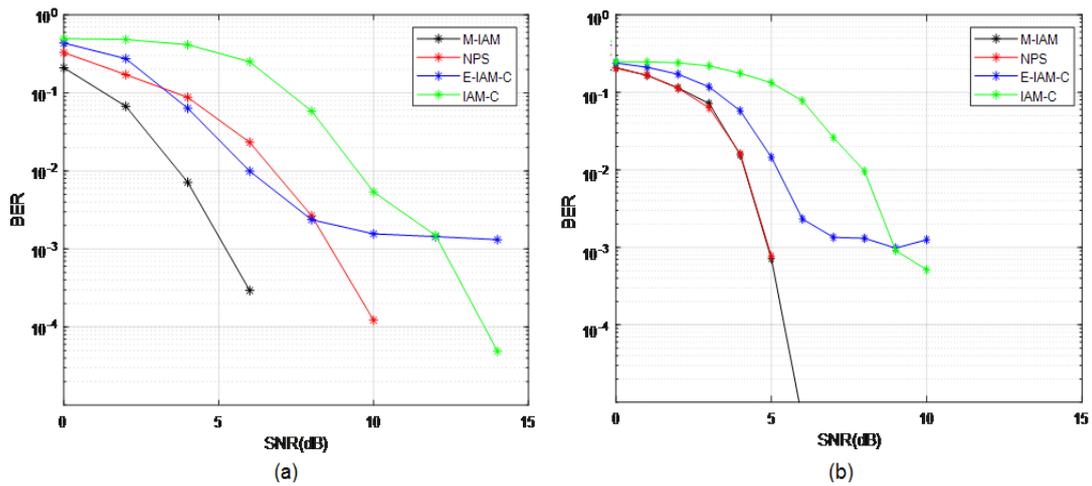

Figure 4. BER Performance for Different Preamble Schemes in IEEE 802.11 Chanel Models for M: (a) 512, (b) 2048 Subcarriers.

International Journal of Computer Networks & Communications (IJCNC) Vol.12, No.1, January 2020

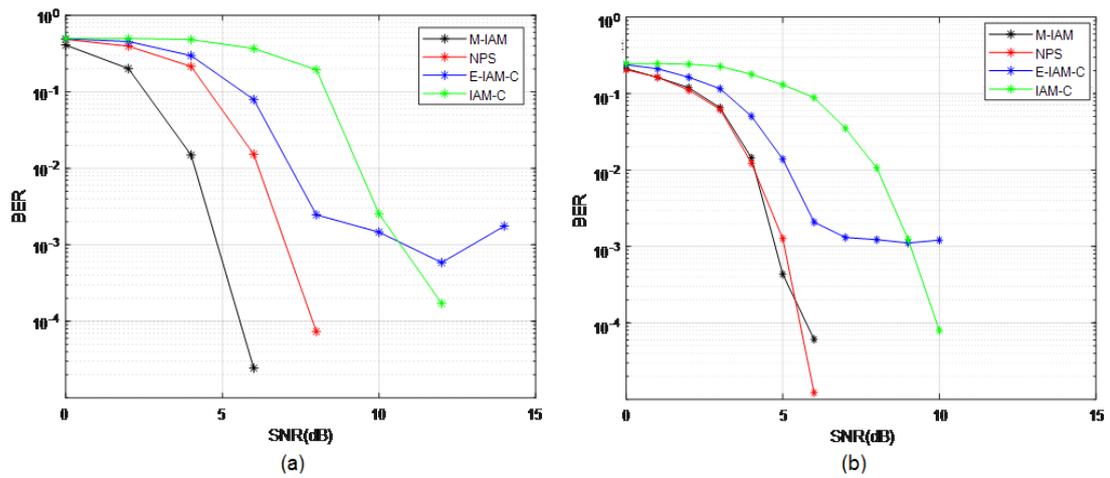

Figure 5. BER Performance for Different Preamble Schemes in Rayleigh Chanel Models for M: (a) 512, (b) 2048 Subcarriers.

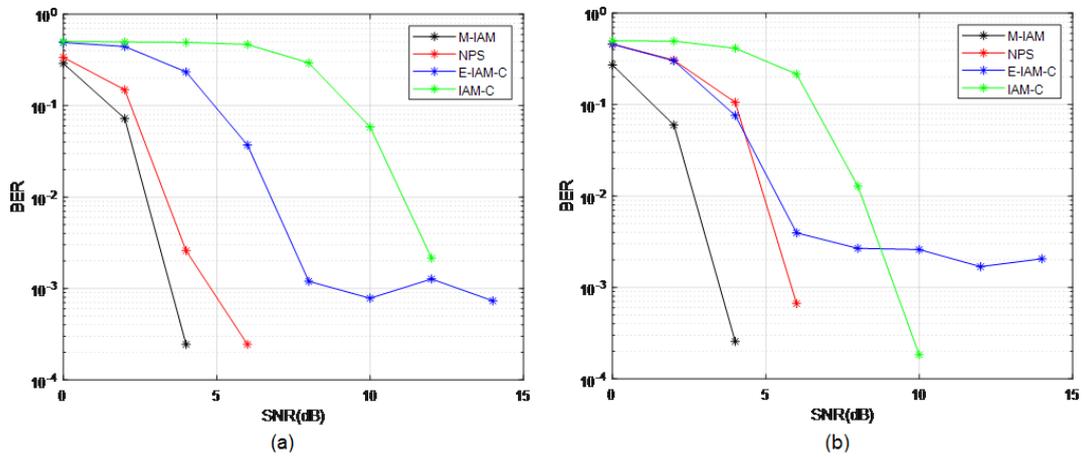

Figure 6. BER Performance for Different Preamble Schemes in Rician Chanel Models for M: (a) 512, (b) 2048 Subcarriers.

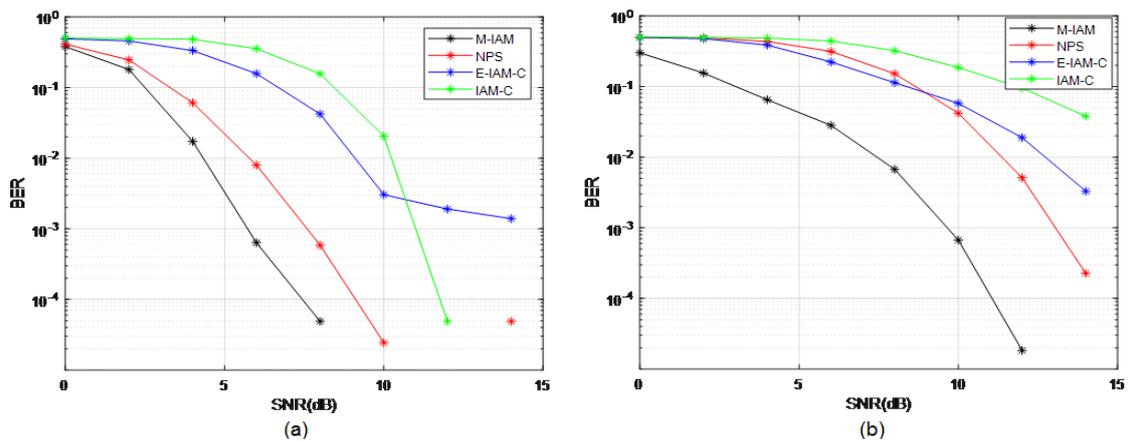

Figure 7. BER Performance for Different Preamble Schemes in Vechular-A Chanel Models for M: (a) 512, (b) 2048 Subcarriers.

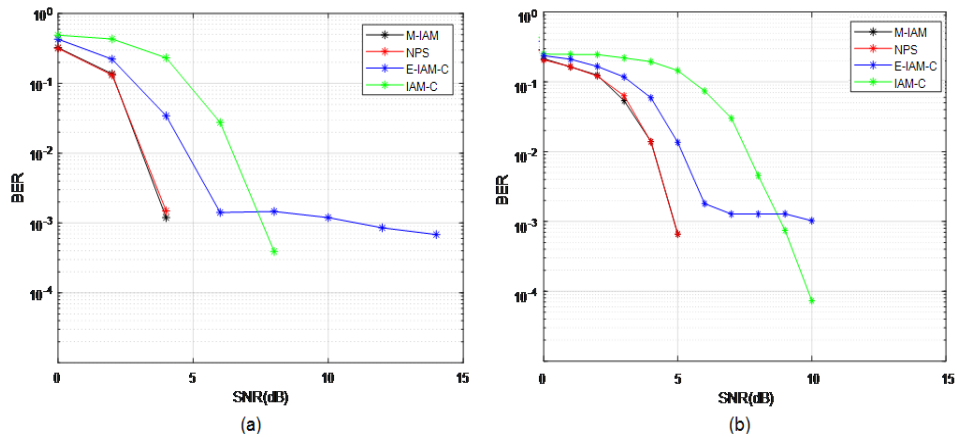

Figure 8. BER Performance for Different Preamble Schemes in AWGN Chanel Models for M: (a) 512, (b) 2048 Subcarriers.

Fig. 9 illustrates the NMSE performance of the different preamble schemes in the IEEE 802.22 channel. It is observed that the performance of the proposed M-IAM has a gain of 1dB over the NPS at NMSE of 10-1 for M=512. However, the NPS has a slightly better performance in the case of M=2048. Fig.10 illustrates the NMSE performance of the various preamble schemes in an IEEE 802.11 channel. It is clear that the proposed M-IAM scheme has a gain of 3 dB over the nearest curve of the other schemes (which is that of NPS) at NMSE of 10-1 for M=512 and 3 dB for M=2048. In Fig.11, the NMSE performance of the various preamble schemes is illustrated in the case of a Rayleigh channel. It is clear that the NMSE performance of the proposed M-IAM is nearly the same as in NPS in case of M=512. However, the proposed M-IAM has a gain of 2 dB at NMSE of 10-1 at M=2048. Fig. 12 illustrates the NMSE performance of the different preamble schemes in the rising channel. It is observed that the proposed M-IAM has a gain 4 dB over the nearest curve of the other schemes (which is that of NPS) at NMSE of 10-1 for M=512. In the case of M=2048, the M-IAM has a gain 3 dB over the NPS scheme. Fig. 13 illustrates the NMSE performance of the different preamble schemes in the Veh-A channel. It is observed that the proposed M-IAM has a gain 0.5 dB over the NPS at NMSE of 10-1 for M=512 and 2 dB in the case of M=2048. It is also observed in fig.9 to fig 13 that the NMSE performance of the M-IAM for M=2048 is better than the NMSE performance of the M-IAM for M=2048. This is because increasing the number of subcarriers results in closer sub channels and hence CFR is totally flat, this means that the assumption mentioned above in (4) is more accurate.

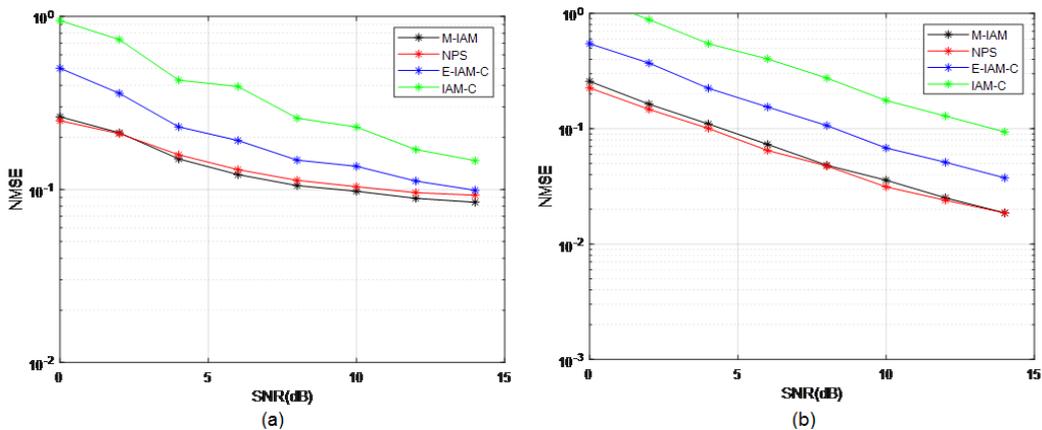

Figure 9. NMSE Performance for Different Preamble Schemes in IEEE 802.22 Chanel Models for M: (a) 512, (b) 2048 Subcarriers.

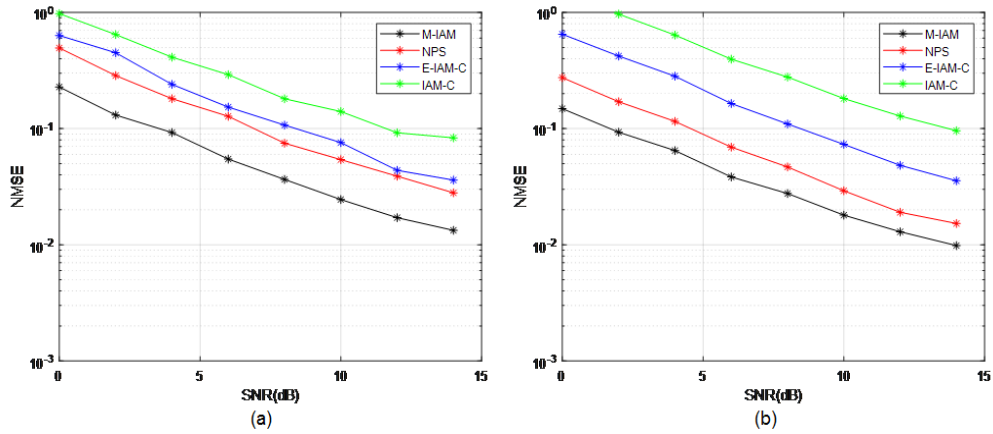

Figure 10. NMSE Performance for Different Preamble Schemes in IEEE 802.11 Chanel Models for M: (a) 512, (b) 2048 Subcarriers.

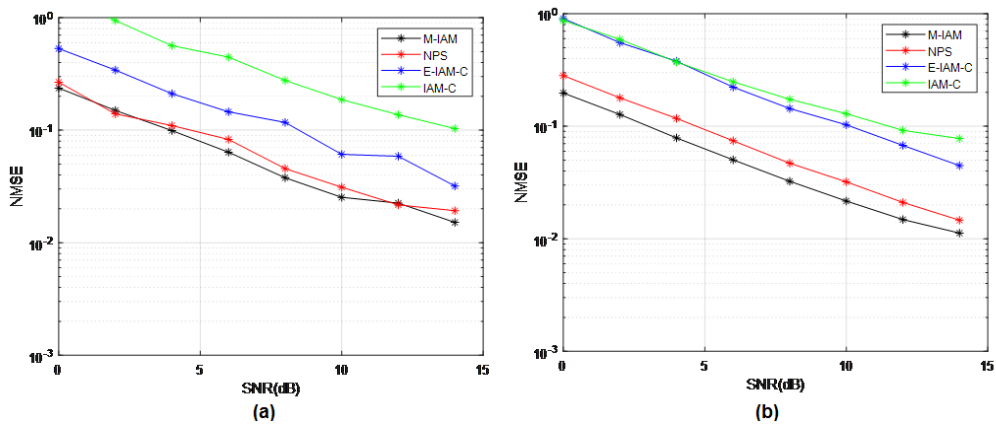

Figure 11. NMSE Performance for Different Preamble Schemes in Rayleigh Chanel Models for M: (a) 512, (b) 2048 Subcarriers.

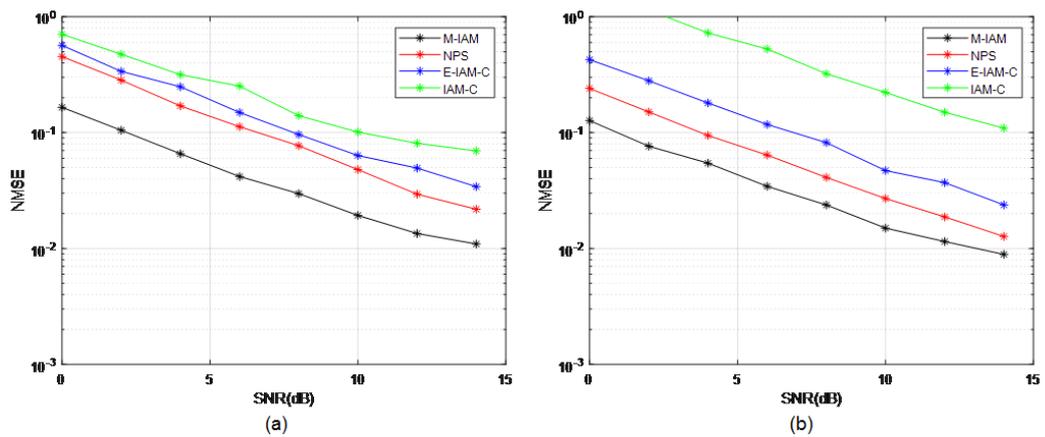

Figure 12. NMSE Performance for Different Preamble Schemes in Rician Chanel Models for M: (a) 512, (b) 2048 Subcarriers.

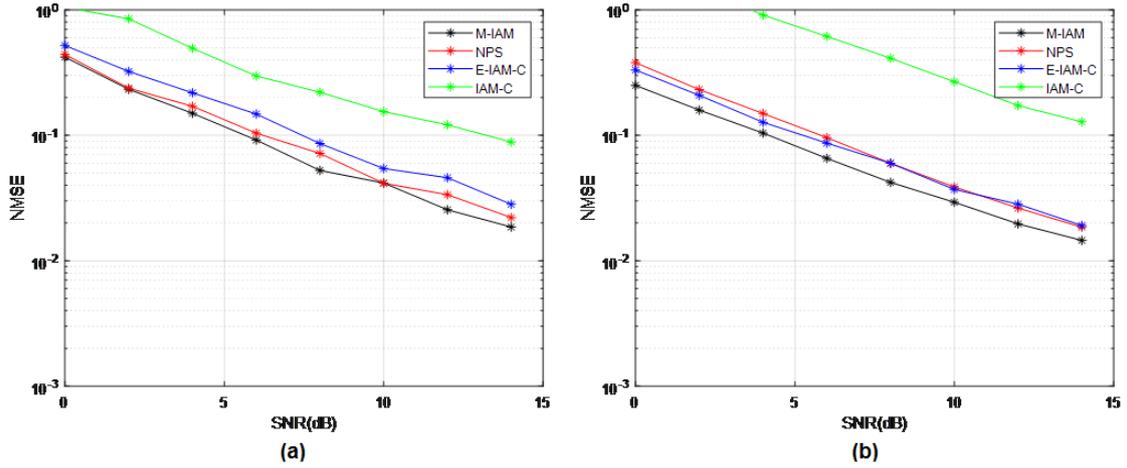

Figure 13. NMSE Performance for Different Preamble Schemes in Vechular-A Chanel Models for M: (a) 512, (b) 2048 Subcarriers.

## 5.3. Peak to Average Power Ratio Comparison

Transmitting data with low power is an important issue in 5G as it results in longer battery life and a reduction in usage energy almost by 90% to support green technology. So in this paper, we focus on reducing PAPR of the M-IAM as high PAPR reduces power amplifier efficiency [21]. PAPR of FBMC system is defined as the ratio of peak power to the average power [22] and is given by:

$$PAPR = \left( \frac{\max\{s(l)^2\}}{E[|s(l)|^2]} \right) \qquad (19)$$

Where $s(l)$ a discrete-time signal of FBMC synthesis is filter bank (SFB) and is given by:

$$s(l) = \sum_{m=0}^{M} \sum_{n} d_{m,n} g_{m,n}(l) \qquad (20)$$

The Performance of PAPR is measured either by the cumulative distribution function (CDF) or by the complementary CDF (CCDF). However, CCDF is more commonly used than the (CDF) [22]-[25]. CCDF is defined as the probability that PAPR exceeds some threshold value that is denoted by (PAPR0). PAPR0 is a random variable (R.V), of which lower values indicate that to better performance

Fig. 14 and Fig. 15 show the PAPR of the proposed M-IAM scheme as compared to other preamble schemes, namely NPS, E-IAM-C and IAM-C for M=512 and M= 2048. It is observed that Fig. 14 and Fig. 15, the proposed M-IAM scheme saves power compared to NPS and E-IAM-C. However, But compared to IAM-C the proposed M-IAM doesn't save power because the IAM-C scheme transmits training symbols at only the middle pilot position but the proposed M-IAM scheme transmits three training symbols at positive odd indexed subcarriers. However, the proposed M-IAM is still better than IAM-C because the latter achieves the worst BER and NMSE results performance over the other preamble schemes as shown by Fig. 3 to Fig. 13 which makes it not valid to be used as a channel estimation scheme in 5G.

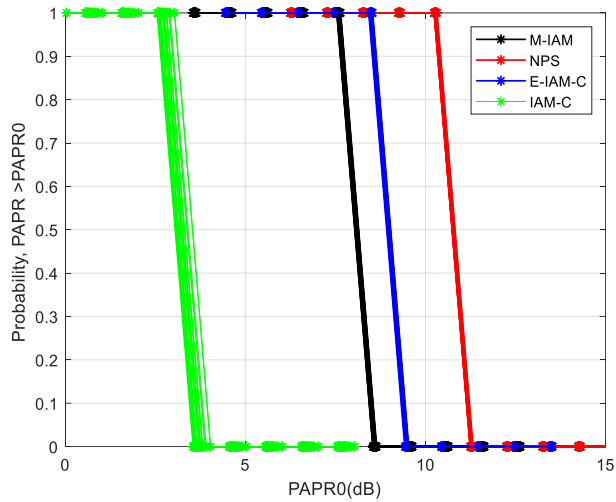

Figure 14. CCDF of Different Preamble Schemes with M=512.

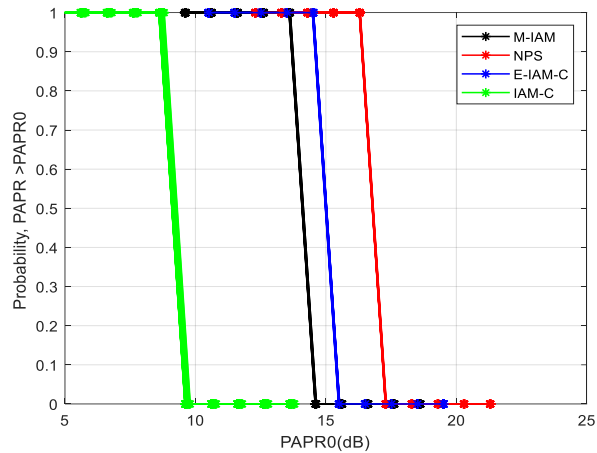

Figure 15. CCDF of Different Preamble Schemes with M=2048.

## 6. CONCLUSIONS

A modified IAM (M-IAM) scheme is proposed to improve channel estimation performance while maintaining lower power consumption. The proposed scheme is characterized by a pseudo pilot magnitude larger than those of most of the conventional preamble schemes. The performance of the proposed M-IAM scheme is illustrated through simulation results and is compared with other conventional preamble schemes in outdoors and indoors environments characterized by time-variant and time-invariant multipath fading channels. The channels selected are the IEEE 802.22 and IEEE 802.11 Rayleigh, Rician, Vehicular-A, and AWGN channels. The results illustrate that the proposed M-IAM scheme has better BER and NMSE performances than those of the conventional schemes. As far as the PAPR is concerned, the results show that the proposed M-IAM has the lowest PAPR over the other preamble schemes except for IAM-C which has lower PAPR than the M-IAM. However, the IAM-C has the worst BER and NMSE performance over the other preamble schemes. It is thus clear that larger pseudo pilot magnitude and lower PAPR are optimized through the M-IAM which makes it more efficient to achieve 5G requirements regarding long battery life and lower power consumption.